\documentclass[aps,showpacs,pra,amssymb, amsmath, twocolumn]{revtex4}
\usepackage{graphicx}
\usepackage{amssymb}
\usepackage{amsmath}
\usepackage{bm}
\usepackage{xcolor} 
\usepackage{array}
\usepackage{multirow}
\usepackage{tabularx}
\usepackage{booktabs}
\usepackage{textcomp}

\newcommand{\bra}[1]{\ensuremath{\langle #1 |}}
\newcommand{\ket}[1]{\ensuremath{| #1 \rangle}}

\begin{document}
\title{Coherent and dissipative dynamics of entangled few-body systems of Rydberg atoms}
\author{Woojun Lee, Minhyuk Kim, Hanlae Jo, Yunheung Song, and Jaewook Ahn}
\address{Department of Physics, KAIST, Daejeon 305-701, Korea}

\begin{abstract}
Experimentally observed quantum few-body dynamics of neutral atoms excited to a Rydberg state are numerically analyzed with Lindblad master equation formalism. For this, up to five rubidium atoms are trapped with optical tweezers, arranged in various two-dimensional configurations, and excited to Rydberg 67S state in the nearest-neighbor blockade regime. Their coherent evolutions are measured with time-varying ground-state projections. The experimental results are analyzed with a model Lindblad equation with the homogeneous and inhomogeneous dampings determined by systematic and statistical error analysis. The coherent evolutions of the entangled systems are successfully reproduced by the resulting model analysis for the experimental results with optimal parameters in consistent with external calibrations.
 \end{abstract}

\maketitle

\section{Introduction}

Neutral atoms have been a promising candidate platform for quantum information science and quantum many-body physics studies~\cite{briegel2000, deutsch2000, saffman2016, gross2017}. They have well-defined energy levels, long coherence and lifetimes, which are all essential for their usage as qubits in quantum information science~\cite{saffman2016}. Furthermore, atoms can be controlled as individual quanta~\cite{schlosser2002, nakagawa2009}, rather than as a collective ensemble, through the developments in laser cooling and trapping techniques. In recent demonstrations, as many as one hundred single atoms were arranged with a set of independently controlled optical tweezers~\cite{nogrette2014, lee2016, kim2016, barredo2016, endres2016, Weiss2018, BarredoNature2018} and entangled through excitation to Rydberg states~\cite{urban2009,isenhower2010,wilk2010, maller2015, jau2016}. With these entangled single-atom systems, Rydberg quantum simulators were constructed, having  about 25-51 qubits, and used to probe the many-body dynamics of Ising-type or XY quantum spin models across phase transitions~\cite{labuhn2016, bernien2017, XY}  and also towards thermalization~\cite{kim2018}. 

Rydberg atoms strongly interact with each other, due to the high polarizability and large-scale dipoles, compared to the ground-state atoms. The giant dipole-dipole interaction among closely-lying Rydberg atoms can shift the resonance of the double excitations out of the range of excitation laser bandwidth, inhibiting the excitations of all other atoms during one is excited. This Rydberg dipole-blockade is of much interest as an effective way to implement  entanglements~\cite{wilk2010, jau2016} and C-NOT gates~\cite{isenhower2010, maller2015} in quantum computation and quantum simulation~\cite{kim2018, labuhn2016, bernien2017, XY, weimer2010}. 

Precise measurements and control of the quantum evolution of these atoms, particularly in their entanglements, are highly important in quantum simulations~\cite{weimer2010}, in which measured system dynamics are used to reproduce and predict the dynamics of other many-body quantum systems. However, system dynamics of an entangled many-body system, which is given as a combination of coherent Hamiltonian and dissipative open-system evolution, are vulnerable to environmental errors. In this paper, we present a numerical analysis of experimentally observed quantum few-body dynamics of Rydberg atoms. We first measure the coherent evolutions of up to five rubidium atoms arranged in various two-dimensional configurations and entangled through Rydberg state excitation, and the measured results are analyzed with a model Lindblad master equation with homogeneous and inhomogeneous dephasings.

The rest of this article is organized as follows: In Sec.~\ref{theory1}, we provide a brief theoretical model description of the quantum dynamics of Rydberg atomic systems. Experimental setup and procedure are described in Sec.~\ref{experiment}, before the result in Sec.~\ref{results}, and possible error sources are discussed in Sec.~\ref{discussions}. A summary is given in Sec.~\ref{conclusion}.

\section{Theoretical description}
\label{theory1}

We consider the dynamics of $N$ atoms arranged in two-dimensional space and interacted with light near-resonant to a Rydberg state. The Hamiltonian, without dephasing taken into account, is given by
\begin{equation}
\label{hamiltonian}
\hat{H}=\sum_{j=1}^N \left\{ \frac{\hbar \Omega e^{i\phi}}{2} \hat{\sigma}_x^{(j)} - \frac{\hbar \Delta}{2} \hat{\sigma}_z^{(j)} \right\} + \sum_{k<l} V_{kl} \hat{n}_k \hat{n}_l,
\end{equation}
where $\hat{\sigma}^{(j)}_x=\ket{1}_{j}\bra{0}_{j}+ \ket{0}_{j}\bra{1}_{j}$ and $\hat{\sigma}^{(j)}_z=\ket{0}_{j}\bra{0}_{j}-
\ket{1}_{j}\bra{1}_{j}$ are the Pauli matrices for pseudo spinors defined with $\ket{0}=\ket{g}$ (the ground state) and $\ket{1}=\ket{R}$ (the Rydberg state), and $\hat{n}_k=\ket{1}_{k}\bra{1}_{k}$ is the excitation number. Also, $\Omega$ (with phase $\phi$) is the Rabi frequency, $\Delta$ is the detuning, and $V_{kl}=-C_6/r_{kl}^6$ is the van der Waals interaction~\cite{beguin2013} between two Rydberg atoms separated by a distance $r_{kl}$. 

\begin{figure}[tbp]
\centering
\includegraphics[width=0.45\textwidth]{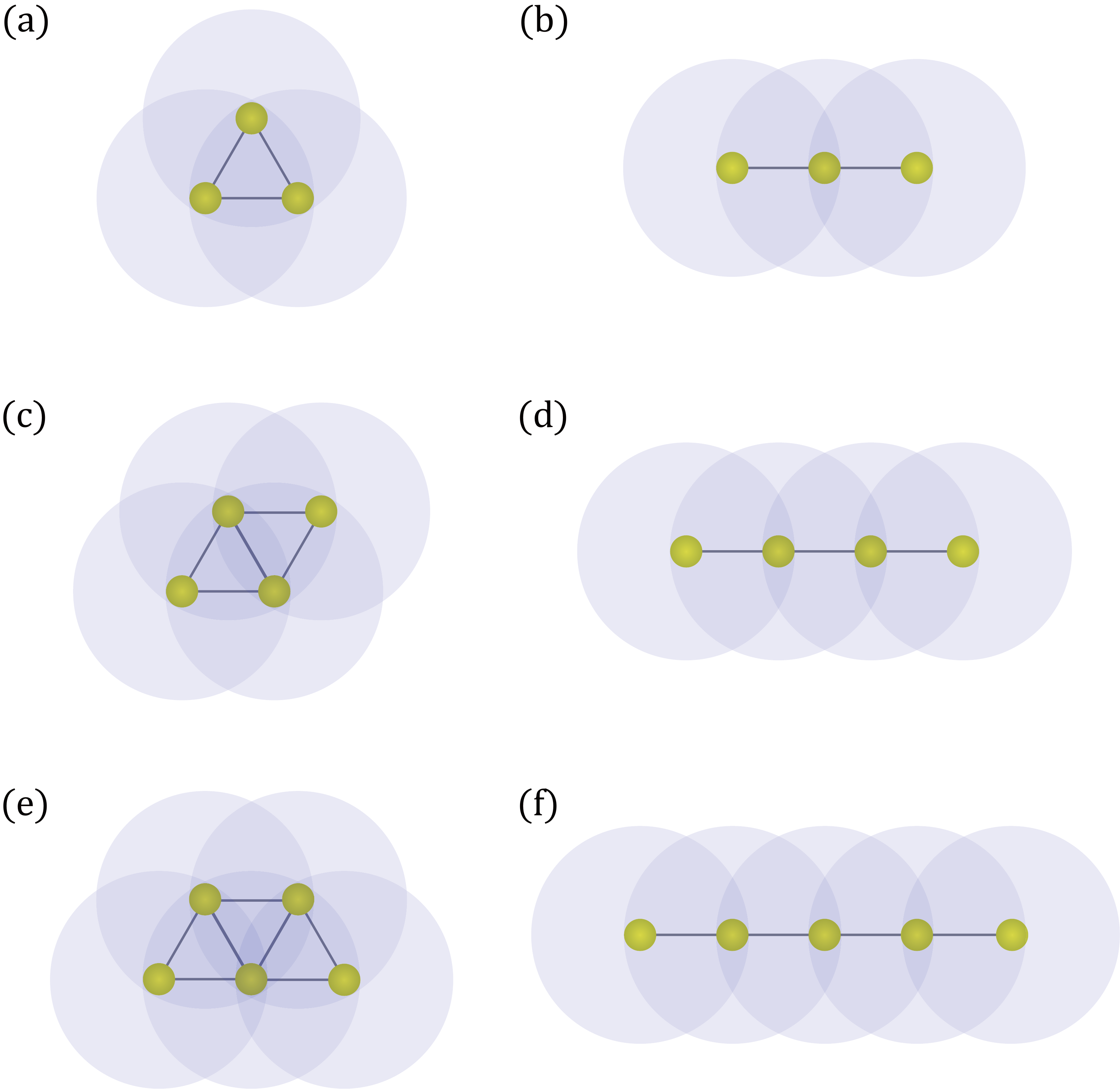}
\caption{(Color online) Atom configurations: (a) triangular three atoms ($r_{12}=r_{23}=r_{13}$); (b) linear three atoms ($r_{12}=r_{23}=r_{13}/2$); (c) zigzag four atoms ($r_{12}=r_{13}=r_{14}=r_{24}=r_{34}=r_{23}/\sqrt{3}$); (d) linear four atoms ($r_{12}=r_{23}=r_{34}=r_{13}/2=r_{24}/2=r_{14}/3$); (e) zigzag five atoms; (f) linear five atoms. Each circle represents the radius of Rydberg blockade, which is larger than the nearest neighbor distance and smaller than the next-nearest neighbor distance, i.e., $r_{(nn)}<r_R<r_{(nnn).}$}
\label{fig1}
\end{figure}

As an exemplary set, we consider six two-dimensional arrangements of $N$=3-5 atoms as shown in Fig.~\ref{fig1}: (a) triangular three ($N$=3) atoms arranged at the vertices of an equilateral triangle, (b) a linear arrangement of three atoms, (c) a zigzag arrangement of four atoms, (d) linear four atoms, (e) zigzag five atoms, and (f) linear five atoms. In the all configurations, the nearest neighbor distance is smaller and the next-nearest is larger than the blockade radius~\cite{urban2009} (i.e., $r_{n.n.}<r_B=(|C_6|/\hbar\Omega)^{-1/6}<r_{n.n.n.}$).
In this case, double excitations of any and only neighboring pairs are prohibited almost, and this prohibition becomes complete in an approximation of ignoring all the longer-distance interactions. Under this approximation, the quantum dynamics of the triangular three atoms in Fig.~\ref{fig1}(a) is a collective Rabi oscillation~\cite{lukin2001}, of which the time evolution is given by 
\begin{equation}
\ket{\psi(t)}=a_0(t)\ket{000}+a_1(t)\frac{\ket{100}+\ket{010}+\ket{001}}{\sqrt{3}},
\label{eq2}
\end{equation}
where $\ket{000}$ is the zero-excitation state and the second term is the superposition of singly-excited states. Likewise, the dynamics of the linear three atoms in Fig.~\ref{fig1}(b) is given in the symmetry basis $\{\ket{000}, \ket{010}, (\ket{100}+\ket{001})/\sqrt{2}, \ket{101}\}$, the zigzag four atoms in Fig.~\ref{fig1}(c) is in $\{\ket{0000}, (\ket{1000}+\ket{0001})/\sqrt{2}, (\ket{0100}+\ket{0010})/\sqrt{2}, \ket{1001}\}$, and so on.

Dephasing of a mixed state is in general described by a Lindblad master equation~\cite{braaten2017, lindblad1976, gorini1976}, which reads:
\begin{equation}
\label{lindblad}
\frac{d\rho}{dt} = -\frac{i}{\hbar} \left[ H,\rho \right]
+\mathcal{L}_{\rm ind}(\rho) + \mathcal{L}_{\rm c}(\rho)
\end{equation}
where $\rho$ is a $2^N$-by-$2^N$ density matrix, $\mathcal{L}_{\rm ind}$ and $\mathcal{L}_{\rm c}$ are the Lindblad superoperators for individual and collective dephasings, respectively, given by
\begin{eqnarray}
\label{individualsuper}
\mathcal{L}_{\rm ind}(\rho) &=& \sum_{j=1}^N \left( L_j \rho L_j^{\dagger} -\frac{1}{2}  \{ L_j^{\dagger} L_j, \rho \} \right) \\
\label{collectivesuper}
\mathcal{L}_{\rm c}(\rho) &=& L_0^{ } \rho L_0^{\dagger} -\frac{1}{2}  \{ L_0^{\dagger} L_0, \rho \}.
\end{eqnarray}
In Eq.~\eqref{individualsuper}, $L_j$ is the Lindblad operator for individual (atom $j$) dephasing, given by
\begin{equation}
\label{Lj}
L_{j}=I^{(1)} \otimes I^{(2)} \cdots \otimes \sqrt{\frac{\gamma_{\rm ind}}{2}}\sigma_z^{(j)}  \cdots \otimes I^{(N)},
\end{equation}
where $I$ is the 2-by-2 identity matrix and $\gamma_{\rm ind}$ is the individual dephasing rate. In Eq.~\eqref{collectivesuper}, $L_{0}$ is the Lindblad operator for collective dephasing, given as a sum of  $L_j$ with collective dephasing rate $\gamma_{\rm c}$ replacing the individual dephasing $\gamma_{\rm ind}$ in Eq.~\eqref{Lj}. As to be explained in Sec.~\ref{results}, in our experiment, the individual dephasing is mainly caused by the spontaneous emission through intermediate state and the collective dephasing is in our experiment is negligible.

Additionally, the phase $\phi$ of the Rabi frequency $\Omega e^{i\phi}$ in Eq.~\eqref{hamiltonian} changes in time, due to the phase noise of Rydberg-state excitation lasers, which induces apparently a dephasing behavior, as recently discussed in a single-body dephasing model~\cite{de leseleuc2018}. In the interaction picture, where the phase is eliminated from the Rabi frequency and treated as a detuning, the Hamiltonian $H'=UHU^\dagger-i\hbar U \dot{U}^\dagger$, basis-transformed with phase-rotation
$U=\Pi_{j=1}^N \left( \ket{0}_{j}\bra{0}_{j}+ e^{i\phi(t)}\ket{1}_{j}\bra{1}_{j} \right)$, is given by
\begin{equation}
\label{hamiltonian2}
\hat{H}'=\sum_{j=1}^N \left\{ \frac{\hbar \Omega}{2} \hat{\sigma}_x^{(j)} - \frac{\hbar (\Delta+\dot{\phi}(t) )}{2} \hat{\sigma}_z^{(j)} \right\} + \sum_{k<l} V_{kl} \hat{n}_k \hat{n}_l,
\end{equation}
where $\Delta(t)=\Delta+\dot{\phi}(t)$ is the time-dependent phase, often analyzed as a Fourier series, i.e.,
\begin{equation} \Delta(t)=  2 \int |\widetilde{\Delta}(f)| \cos[2\pi ft+\xi(f)] df, 
\label{df}
\end{equation}
where $|\widetilde{\Delta}(f)|$ and $\xi(f)$ are the spectral amplitude and phase of $\Delta(t)$. So, the laser phase noise in repetitive measurements randomizes $\xi(f)$ and induces $\Delta(t)$ fluctuations, which causes an apparent dephasing behavior in the given quantum dynamics.

\section{Experimental setup and procedure}
\label{experiment}

\label{setup}
\begin{figure*}[!thb]
\centering
\includegraphics[width=0.85\textwidth]{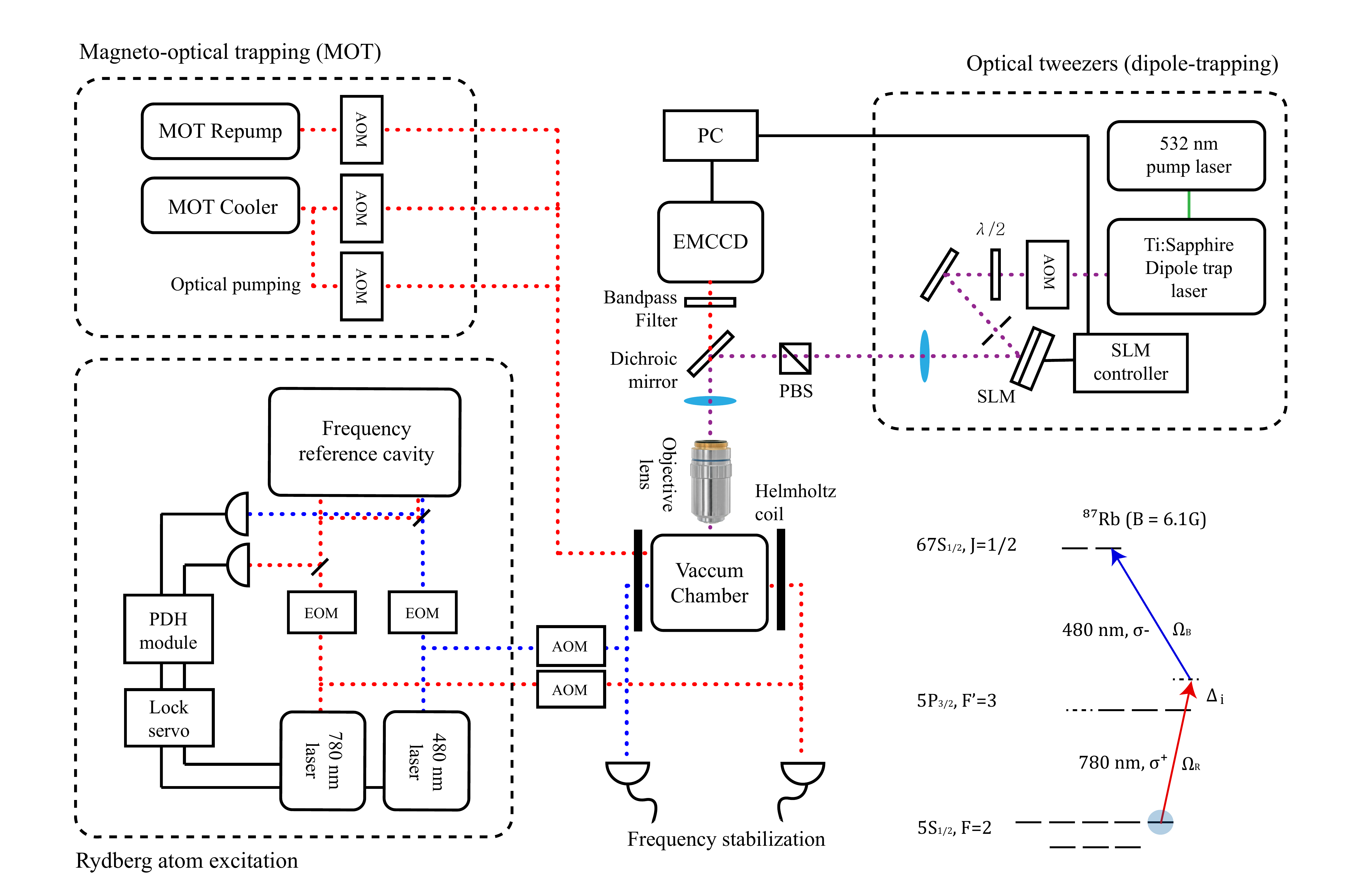}
\caption{(Color online) A schematic diagram of the experimental setup and the energy level diagram for Rydberg-state excitation (AOM:acousto-optic modulator, EOM:electro-optic modulator, PC:personal computer, SLM:spatial light modulator, PBS: polarization beam splitter)}
\label{fig2}
\end{figure*}

The experimental setup is shown in Fig.~\ref{fig2}(a), which is similar to our earlier reports~\cite{lee2016, kim2016,lee2017, kim2018}. In brief, the setup consists of a magneto-optical trap for cold rubidium atoms ($^{87}$Rb), a control system of optical tweezers (far-off resonance dipole traps), and an optical system for Rydberg excitation. Rubidium atoms were first cooled  to 30~$\mu$K through Doppler and polarization-gradient coolings. During the cooling stage, optical tweezers (of 820-nm wavelength, 1-mK trap-depth, and 1.4-$\mu$m diameter) were trapping atoms in pre-determined target sites and the MOT was turned off by shutting off the anti-Helmholtz coils. Typical arrays before rearrangements were about half-filled, due to collisional blockade~\cite{schlosser2002}. So, the occupancy or vacancy in each optical tweezer was checked with fluorescence imaging, $\ket{5S_{1/2}, F=2} \rightarrow \ket{5P_{3/2}, F'=3}$, by an electron multiplying charge-coupled device (EMCCD). After the occupancy was all checked, unity-filled arrays were then created with reconfiguration of captured atoms~\cite{lee2017, barredo2016} through two times of the three-step processes of imaging, vacancy-filling, and verification. The six different atom configurations, introduced in Sec.~\ref{theory1}, were tested, which were linear or zigzag $N=3$, 4, or 5 atoms. 

After a unity-filled atom arrangement was prepared, Rydberg-state excitation was performed with a two-photon transition from $\ket{g}=\ket{5S_{1/2}, F=2, m_F=2}$ to $\ket{R}=\ket{67S_{1/2}, m_J=1/2}$ via off-resonant intermediate state $\ket{m}=\ket{5P_{3/2}, F'=3, m_F'=3}$. We used 780-nm and 480-nm lasers (diode lasers from Toptica), counter-propagating with $\sigma^+$ and $\sigma^{-}$ polarizations, respectively. The Rabi frequency of the two-photon transition is given by $\Omega=\Omega_{780}\Omega_{480}/(2\Delta_i)=(2\pi)$1.0~MHz, where $\Omega_{780}=(2\pi)94$~MHz and $\Omega_{480}=(2\pi)12$~MHz are the Rabi frequencies of the one-photon transitions ($\ket{g}\rightarrow \ket{m}$ and $\ket{m}\rightarrow \ket{R}$) and $\Delta_i =-(2\pi) 560$~MHz is the one-photon detuning of the 780-nm laser from the intermediate transition ($\ket{g}\rightarrow \ket{m}$). 
The phase of the Rabi frequency is $\phi=\phi_{780}+\phi_{480}$, the sum of the phases of the lasers. The frequencies of the lasers were stabilized to a narrow linewidth of $<$$(2\pi)30$~kHz with an ultralow expansion (ULE) reference cavity (from Stable Laser Systems). The ULE cavity had a finesse of 15,000 and AR coated at dual wavelengths of 780~nm and 480~nm. The laser wavelengths were roughly monitored by a wavemeter (HighFinesse WS7-60) within 60~MHz accuracy and Pound-Drever-Hall locking technique (PDH module from Stable Laser Systems and PDD110 from Toptica) was adopted to lock the laser frequencies to the Fabry-Perot signal reflected from the reference cavity, in conjunction with fast lock servos (FALC110 from Toptica). The stray E-field was suppressed by grounded electrodes placed around the chamber.

\begin{figure}[!h]
\centering
\includegraphics[width=0.48\textwidth]{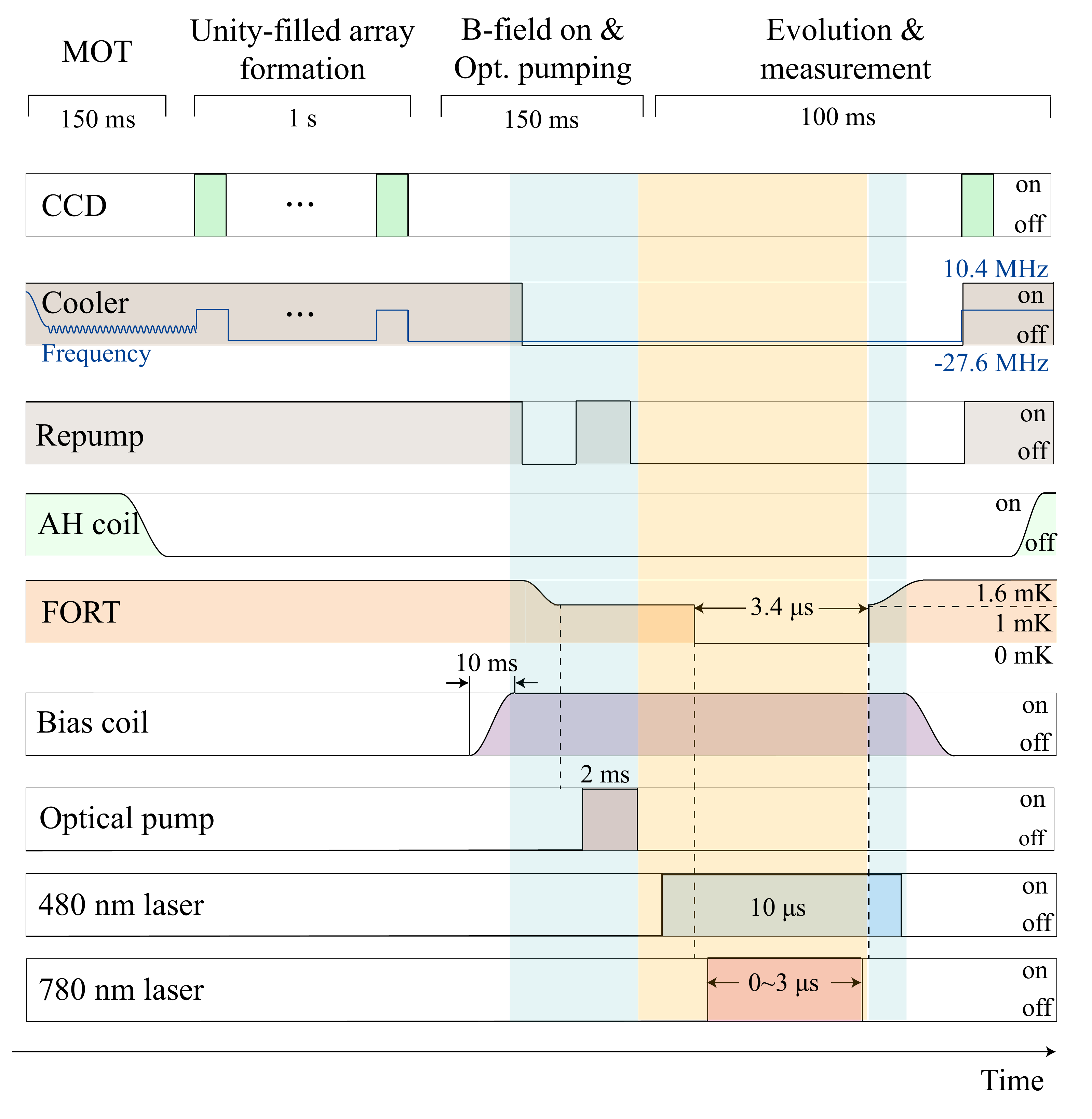}
\caption{(Color online) Experimental procedure: the time sequences of fluorescence imaging, MOT cooling and repumping, anti-Helmholtz current, far-off resonant trapping (opitcal tweezers), bias B-filed, optical pumping, and Rydberg-state excitation with 480-nm and 780-nm lasers, respectively.}
\label{fig3}
\end{figure}

The time sequence of the experimental procedure is summarized in Fig.~\ref{fig3}. Before the Rydberg-state excitation, optical pumping to $\ket{g}$ was performed for 2~ms, when the quantization axis was defined with a Helmholtz bias coil ($B=6.1$~G). Then, we turned on the 480-nm laser, turned off the optical tweezers for 3.4~$\mu$s to avoid push-out of atoms in the Rydberg states (due to the light-induced potential), and finally turned on the 780-nm laser for Rydberg-state excitation. After the 780-nm laser turn-on with various pulse durations, the optical tweezers were turned back on to recapture the atoms in $\ket{g}$. Whether each atom was recaptured or not (a projection measurement to $\ket{g}$) was recorded with the fluorescence imaging through $\ket{5P_{3/2}, F'=3}$.
\label{setup}

\section{Results}
\label{results}
\begin{figure*}[!htbp]
\centering
\includegraphics[width=\textwidth]{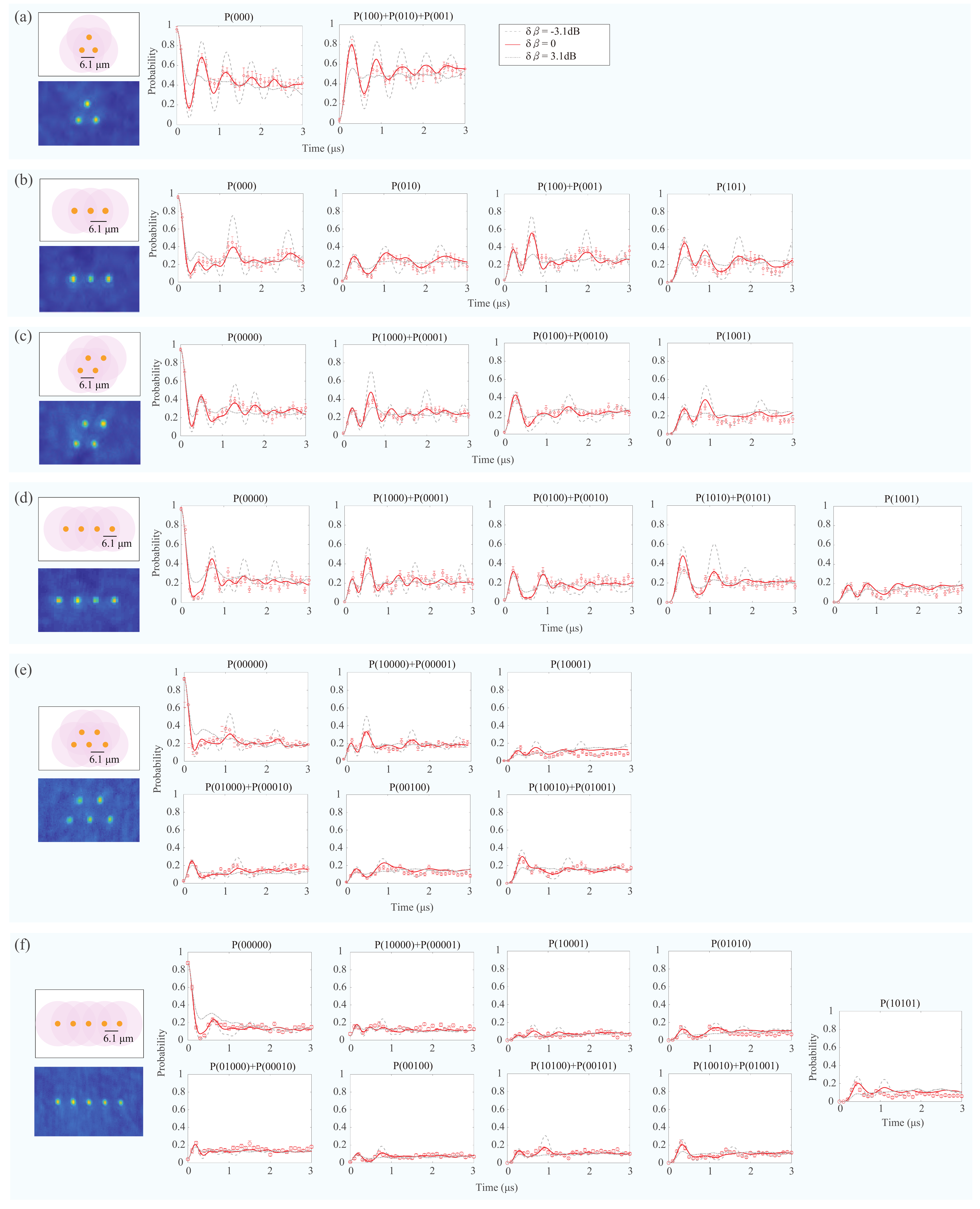}
\caption{(Color online) (a) Triangular three ($N=3$) atoms: geometry and image in the leftmost column, respectively, and the experimental measured probability (data points) compared with the numerical calculations for optimal fitting (solid line) and $\pm 3.1$dB shifts in $\beta$ (dot-dashed and dashed lines), for each symmetric basis, where ``0'' and ``1'' in the states indicate $\ket{g}$ and $\ket{R}$ states, respectively. Same for the other configurations: (b) linear $N=3$ atoms, (c) zigzag $N=4$ atoms, (d)  linear $N=4$ atoms, (e) zigzag $N=5$ atoms, and (f) linear $N=5$ atoms.}
\label{fig4}
\end{figure*}

Experimentally measured quantum dynamics are summarized in Fig.~\ref{fig4}. The results for the total six configurations, $N$=3-5 atoms in linear or zigzag configuration, are shown in Fig.~\ref{fig4}(a)-(f), where the Rydberg blockade radius was $r_R= 8.8(3)$~$\mu$m and the lattice constant was $d=6.1(3)$~$\mu$m. For example, the case of the equilateral triangular three atoms is shown in Fig.~\ref{fig4}(a), where the schematic geometry and the image of the atoms are shown in the leftmost column, and the measured probabilities are in the right columns. The time-evolving state probabilities are plotted for the symmetry bases $\ket{000}$ and $(\ket{100}+\ket{010}+\ket{001})/\sqrt{3}$, as in Eq.~\eqref{eq2}. The scan range of the quantum evolution was $0-3$~$\mu$s with a time step of 0.1~$\mu$s (total 31 data points with 150 repetitive measurements). Similarly, the three atoms in the linear configuration is shown in Fig.~\ref{fig4}(b), where the constituent symmetry bases are $\ket{000}$, $\ket{010}$, $(\ket{100}+\ket{001})/\sqrt{2}$, and $\ket{101}$. The remaining configurations are also represented, with probability measurements for the corresponding sets of symmetry bases.

In comparison, numerical calculations was performed with Eqs.~\eqref{lindblad}~and~\eqref{hamiltonian2}, taking into account the contributions of experimental and measurement errors. As to be discussed in Sec.~\ref{discussions}, major error sources are: (a) the spontaneous decay from $\ket{R}$ to $\ket{g}$,  (b) optical-tweezer atom loss, (c) leakage to intermediate state, and (d) laser noise. Table~\ref{table1} summarizes the experimental uncertainties related to these error sources. Noises in laser intensity and phase were $\delta I/I=3$\% and $|\Delta|=0.4\Omega$, respectively. The position uncertainty of the optical tweezers was $\delta r/r=5$\%. The measurement uncertainty was $\delta P=3$\%, mainly from the spontaneous emission from $\ket{R}$ to $\ket{g}$ and also atom escapes from optical tweezers. 

\begin{table}[!htbp]
\centering
\caption{Error sources and amounts}
\label{table1}
\renewcommand{\arraystretch}{1.2} 
\begin{tabular}{l r}
\hline\hline
Error sources & \\ 
\hline
Projection measurement error $\delta P(g|R)$ & $\delta P=3$\% \\ 
$\cdot$ Spontaneous emission $\ket{R}\rightarrow \ket{g}$ & $\sim$3\% \\ 
$\cdot$ Background atom entering optical tweezers & $\sim$0.01\% \\ 
$\cdot$ De-excitation to other ground states & $\sim$0.1\% \\ 
\hline
Projection measurement error $\delta P(R|g)$ & $\delta P=3$\% \\ 
$\cdot$ Atom escaping from optical tweezers & $\sim$3\% \\ 
$\cdot$ Background atom collision & $\sim$1\% \\ 
\hline
Dephasing rate & $\gamma =1$~MHz \\ 
$\cdot$ Leakage to intermediate state & $60$~kHz \\
$\cdot$ Rydberg-excitation laser intensity noise & $\gamma_{\rm c}\ll 30$~kHz \\ 
$\cdot$ Rydberg-excitation laser phase noise & $\sim$1~MHz \\ 
\hline\hline
\end{tabular}
\end{table}

Each solid line in Fig.~\ref{fig4} represents the result of the calculation. We used
two parameter fitting with $\alpha \equiv \Omega/\Omega_0$ and $\beta \equiv |\widetilde{\Delta}(f)|/|\widetilde{\Delta}(f)|_0$, where $\Omega_{0}$ and $|\widetilde{\Delta}(f)|_{0}$ are the references retrieved from single-atom experiments. After randomization with $\xi(f)$ in Eq.~\eqref{df}, we obtain $(\alpha, \beta) =$ (0.94, 3.1~dB), (1.04, 0.0~dB),  (0.96, 0.0~dB), (1.02, 1.3~dB), (0.96, 0.6~dB), and (0.96, 3.1~dB) for the six configurations, respectively. The dashed and dot-dashed lines in each figure are the calculations with $\delta\beta=\pm3$~dBm shifts, respectively, from the above values. To estimate how well the  experimental data are replicated by the model fitting, $R^2$ values are calculated, which are the proportion of the measured behaviors explained by the model. With the optimal fitting conditions of ($\alpha$, $\beta$), we get $R^2=0.90(1)$ (e.g., for $N=3$ cases) and this value gradually decreases below $0.6$ when $\Delta\alpha=20$\% or $\Delta\beta=\pm 3$~dB. Note that calculations without the phase noise taken into account give a similar $R^2$ values below $0.6$.

\section{Discussions}
\label{discussions}
Deviations from ideal dynamics, for example, a simple two-state oscillation for the triangular three atoms in Fig.~\ref{fig3}(a), are attributed to mainly four different physical reasons: (a) sources of projection measurement error $P(R|g)$, (b) sources of projection measurement error $P(R|g)$, (c) dephasing due to leakage to intermediate state, and (d) laser noises. In the following, these error sources are discussed.

(a) The measurement error $P(g|R)$, the conditional probability of false measurement of $\ket{g}$ given that the state was initially in $\ket{R}$, mainly comes from the spontaneous emission from $\ket{R}$ to $\ket{g}$. 
A certain portion of Rydberg atoms, that are supposed to be absent at measurements, can be found trapped due to the spontaneous decay before escaping. Using the Rydberg atom lifetime $\tau_R=140$~$\mu$s, our numerical simulation gives an estimated probability $P(g|R)$ of around 3\%. Therefore, for example, the theoretical probability (solid lines) in Fig.~\ref{fig4} of three atoms in $\ket{RRR}$ is calculated as $P'_{RRR}=\{1-P(g|R)\}^{3} P_{RRR}$ to compare with the actual (false) measurement (data points) in Fig.~\ref{fig4}. In addition, there are minor sources of contribution to $P(g|R)$: background atoms can enter the trap and increase $P(g|R)$, with negligibly small probability of $<$0.01\%; coherent de-excitation from $\ket{R}$ can evolve to other ground hyperfine states, due to imperfect laser polarizations, which results in false $\ket{g}$ measurement of about 0.1\%.

(b) The measurement error $P(R|g)$, the conditional probability of false measurement of $\ket{R}$ given that the state was initially in $\ket{g}$, is mainly caused by atoms escaping from optical tweezers, which is typically $<3$\% in our experiments. In addition, the atom escape probability due to background collision causes false $\ket{R}$ measurement in our experiment, estimated $<1$\%.

(c) The transition probability to the intermediate state ($\ket{5P_{3/2}}$) is nonzero during laser excitation, and the fast spontaneous decay from $\ket{5P_{3/2}}$ to $\ket{g}$ results in non-Hermitian dynamics. The leakage estimated from the detuned Rabi oscillation $\Omega_{780}^2 / (\Delta_i^2 + \Omega_{780}^2) \sim 0.2\%$ is small; however, the two-level approximation~\cite{de leseleuc2018} of the given three-level system dynamics gives the individual dephasing rate $\gamma_{\rm ind}$ in Eq.~\eqref{Lj}, which is estimated as $\sim$$(2\pi)20$~kHz. 

(d) Rydberg excitation lasers (780 nm and 480 nm) have intensity and phase noises. The intensity noise is from the laser diode itself, AOM modulation error, and beam pointing error, which results in fluctuations of $\Omega$ and $\Delta$ by AC Stark shift. This fluctuation is measured as about 3\% without feedback (0.7\% with feedback) in our experiment, making little change in measured single-atom dynamics. The phase noise $|\widetilde{\Delta}(f)|$ in Eq.~\eqref{df}, however, is estimated up to with a scaling factor, using the frequency error signals from PDH locking electronics of the lasers. All the analyses in Sec.~\ref{results} are consistent with the single-atom results up to a 3~dB scaling.

\section{Conclusion}
\label{conclusion}
We have presented a numerical model analysis of the experimentally measured quantum few-body dynamics of Rydberg atoms. Using up to five $^{87}$Rb atoms arranged in linear or zigzag configurations and excited to the Rydberg 67S state, we measured the time-evolving probabilities of the atomic systems in all symmetry basis. As a theoretical model, we used the Lindblad master equation for Rydberg atom chains in consideration of experimental error sources, such as the projection measurement errors, the leakage to an intermediate state, and the phase noise of the excitation lasers. The resulting calculation agreed well with the experimentally observed measurements, suggesting that
the quantum dynamics of Rydberg-atom systems are suitably described with the current model extending the single-body dephasing model~\cite{de leseleuc2018} to a few-body problem.

\begin{acknowledgements}
This research was supported by Samsung Science and Technology Foundation [SSTF-BA1301-12], National Research Foundation of Korea (NRF) (2017R1E1A1A01074307), and Institute for Information \& communications Technology Promotion (IITP-2018-2018-0-01402).

\end{acknowledgements}

\end{document}